\documentclass[a4paper,showkeys,floatfix,aps,pre,longbibliography,superscriptaddress,reprint]{revtex4-1}

\usepackage[dvipsnames]{xcolor}
\definecolor{linkcolor}{rgb}{0.3,0.3,1.0} 
\usepackage[pdftex,colorlinks=true, linkcolor= linkcolor, citecolor= linkcolor, urlcolor= linkcolor, hyperindex=true,hyperfigures=true]{hyperref} 

\usepackage{amsmath,amssymb}
\usepackage{graphics,graphicx}
\usepackage{dcolumn,bm}
\usepackage{color}
\begin{document}

\title{Inequality of avalanche sizes in models of fracture}


\author{Diksha}
\affiliation{Department of Physics, SRM University - AP, Andhra Pradesh 522240, India}

\author{Sumanta Kundu}
\affiliation{Department of Physics and Astronomy, University of Padova, 
Via Marzolo 8, I-35131 Padova, Italy}
\affiliation{INFN, Sezione di Padova, Via Marzolo 8, I-35131 Padova, Italy}

\author{Bikas K. Chakrabarti}
\affiliation{Saha Institute of Nuclear Physics, 1/AF Bidhannagar, Kolkata 700064, India}

\author{Soumyajyoti Biswas}
\affiliation{Department of Physics, SRM University - AP, Andhra Pradesh 522240, India}



\begin{abstract}
Prediction of an imminent catastrophic event in a driven disordered system is of paramount importance – from the laboratory scale controlled fracture experiment to the largest scale of mechanical failure i.e., earthquakes. It has been long conjectured that the statistical regularities in the energy emission time series mirrors the “health” of such driven systems and hence have
the potential for forecasting imminent catastrophe. Among other statistical regularities, a measure of how unequal the avalanche sizes are, is potentially a crucial indicator of imminent failure. The inequalities of avalanche sizes are quantified using inequality indices traditionally used in socio-economic systems: the Gini index ($g$), the Hirsch index ($h$) and the Kolkata index ($k$). It is then shown analytically (for mean field) and numerically (for non mean field) in models of quasi-brittle materials that the indices show universal behavior near the breaking points in such models and hence could serve as indicators of imminent breakdown of stressed disordered systems.
\end{abstract}


\maketitle
\section{Introduction}
When a solid is slowly loaded until its breaking point, it goes through many microscopic damages. Such microscopic damages can grow with additional load applied, interact via the stress field modification, and accumulate into a macroscopic failure (for example, a shear plane in the case of compression induced damage in quasi-brittle objects). However, all breaking events are not equally damaging. Indeed, if we go by the acoustic energy signal ($S$) emitted in each event, their size distribution ($P(S)$) usually has a fat-tail ($P(S)\sim S^{-\delta}$), implying a very few events accounting for most of the damages incurred \cite{wiley_book,BONAMY20111}. This `damage inequality' is reminiscent of inequalities observed in a myriad of complex systems, including physical (from crackling noise \cite{crackling} in magnetic domain wall movements to the largest mechanical failures in earthquakes \cite{earthquake}), sociological \cite{Chatterjee_2015}(deaths in armed conflicts, citations of papers) and economic (wealth distribution) \cite{chakrabarti2013econophysics} systems. This fat-tail or power-law behavior is often attributed to the vicinity of a critical point in the system that is reached either through the fine tuning of an external parameter or the system self-organizes to the critical point through a slow drive \cite{PhysRevLett.59.381}. 

It was empirically noted by Pareto in 1896 in his famous 80-20 law that 80\% of wealth is accumulated with 20\% of the richest people. Since then this observation was much broadly applied to see that in many socio-economic systems 80\% of `successes' come from 20\% of `attempts' \cite{Pareto}. Since then, much has been achieved in studies of inequality, primarily in socio-economic systems, mostly because of the immense adverse effect such inequalities could have in socio-economic contexts \cite{inequality}. A much recent interest is to understand the inequality in the responses of physical systems, particularly when such systems are near a critical point \cite{k_pre}. The main goal in such a scenario is to forecast imminent catastrophe (or distance to the critical point) in such systems. For example, it was shown that in models of fracture, the growing inequality of avalanche sizes shows universal behavior near the critical point \cite{k_pre}. It was then shown that in a supervised machine learning algorithm, measures of inequality of avalanche sizes were the most important attribute in predicting the imminent failure point \cite{PhysRevE.106.025003} in models of quasi-brittle materials. It was also shown subsequently that a family of self-organized critical (SOC) systems (e.g., Bak-Tang-Wiesenfeld (BTW) and Manna sandpile models among others) also show universal behavior in terms of the inequality measures of their responses (avalanches) \cite{soc_fbm}. Along that line, it was shown using extensive data analysis in a wide variety of socio-economic systems that have been conjectured to be in self-organized critical states, that the measures of inequalities in such systems show the similar behavior as seen in the SOC models. Such observations underline the importance of the inequality measures of responses of a system near its critical point for uncovering the formal link between such measures and critical phenomena \cite{10.3389/fphy.2022.990278} and also to make use of such a link for the practical purposes of forecasting an imminent critical point in systems where such a point could have a catastrophic consequence (for example, in fractures and earthquakes). 

In this work, we study two generic models of fracture in disordered quasi-brittle solids -- the fiber bundle model (FBM) and the random fuse model (RFM), in terms of the inequality of their response functions (avalanche time series) as they are stressed to their breaking points. Particularly, in the mean-field FBM, we analytically calculate the inequality indices and their scaling behavior near the critical point, thereby illustrating the previously numerically observed universal behavior. We also study the more realistic non-mean field limit of the FBM and the RFM and show numerically that similar scalings are observed (with different exponent values) in such cases as well. 

The paper is organized as follows: first, we define the indices that were used to quantify the inequalities in the avalanche statistics of these models, then we describe the models and their simulations. We then go on to present the analytical calculations for the mean-field case and compare them with previous numerical studies. Then we give the numerical results for the non-mean-field case of the FBM and the RFM. Finally, we discuss the results and their implications for forecasting imminent failure in these models and potentially in experiments and conclude.

\section{Quantification of inequality of avalanches: The inequality indices} 

The inequalities in socio-economic contexts have been quantified using several indices, for example, the Gini-index $g$ \cite{gini} (for wealth), the Hirsch-index $h$ (for citations) \cite{h-index}, and more recently the Kolkata index $k$ \cite{kolkata}, to name a few. The role of these indices in parametrizing socio-economic inequality has expanded over the past few decades. 
 These inequality measures are defined through the Lorenz curve. When a time series is arranged in ascending order of the size of the values in the series, the Lorenz function $L(p)$ gives the cumulative fraction of the total `mass' acquired by the $p$ fraction of the smallest events. If all the values are equal in size, then the Lorenz function would be a diagonal line from the origin to (1,1), known as the equality line. But in general, these values are not equal in size, so the Lorenz function is nonlinear, always staying under the equality line and increasing monotonically from $L(p=0)=0$ to $L(p=1)=1$. The Gini index is then given by twice the area between the Lorenz curve and the equality line, where $g = 0$ implies perfect equality and $g = 1$ shows the extreme inequality. The crossing point of the opposite diagonal line with the Lorenz curve gives the value of the Kolkata index $k$, which says that $(1-k)$ fraction of the largest events account for the $k$ fraction of the total events, where $k = 0.5$ shows the perfect equality and $k = 1$ shows the extreme inequality. The Hirsch-index is generally calculated in the case of citations of individual scientists or scholars but the definition could be expanded to measure the $h$ index of any series. The $h$-index is then the highest number $h$ such that $h$ events have at least a size $h$ each. For a more detailed discussion on the definitions of these measures, see Ref. \cite{10.3389/fphy.2022.990278}.

 In this work, we will be measuring the inequality of the time series of the avalanches recorded for slowly loaded disordered materials, represented by generic threshold activated models. Therefore, for our purposes, the Lorenz function $L(p)$ would represent the fraction of the breaking events resulting from the smallest $p$ fraction of the avalanches. The other definitions would follow the same line. We will show how the inequalities grow as the system gradually approach the catastrophic failure points. Most importantly, we will monitor the values: $g_f, k_f, h_f$ the inequality indices take just prior to the catastrophic breakdown. We will be interested in the universality of these values and also the finite size and off-critical scaling behavior these measures show, particularly for the context relevant to forecasting an imminent failure.

\section{Models and Methods}
As mentioned before, in this work we consider two generic models of fracture -- the fiber bundle model (FBM) \cite{pierce,rmp} and the random fuse model (RFM) \cite{Herrmann,Kahng,Zapperi1997}. In the FBM, an ensemble of $N$ linear elastic fibers are considered, which have different individual breaking thresholds. When a load is applied on all the fibers, some of the weak fibers fail and the load carried by those fibers is redistributed among the remaining ones, which might in-turn fail, leading to an avalanche event. At a critical value of the load, the entire system collapses. In the RFM, an ensemble of electrical fuses are arranged in a network (say, a square lattice). Each fuse has a failure threshold, beyond which it is burnt. When a voltage difference is applied across the system (say, the two ends of a square lattice), the weaker fuses are burnt, requiring the current to be diverted through other paths, which might in-turn cause other fuse burning, leading to an avalanche event. The modification in the flow of current due to burnt fuses mimics the modification of the stress field in an elastic solid due to local damages. Both of these models are very well studied in the literature from various viewpoints of applications in fracture of disordered systems \cite{doi:10.1080/00018730300741518,alex_book} and mimic many of the experimentally observed features such as power-law scaling of avalanches (see e.g. \cite{BONAMY20111}), non-linear stress-strain relations (see e.g. \cite{metal}) etc. 

Finding an indication of the imminent failure in these models, without using detailed information regarding the individual failure thresholds, therefore, is a long-standing crucial issue. This problem has been approached through various different ways, including changes in the avalanche size distribution exponent prior to failure \cite{PhysRevLett.95.125501}, non-monotonic behavior of the elastic energy stored in the system \cite{10.3389/fphy.2019.00106} and so on. However, here we will study the behavior of the inequality indices, which have shown promising indications as precursory measures of imminent breakdown \cite{PhysRevE.106.025003}.

\subsection{Simulation methods}
While a mean-field limit of the FBM is analytically tractable, the more realistic versions of FBM and also the RFM are not. Those limits are accessed through numerical simulations. Therefore, here we describe the simulation methods for the two models using which the avalanche time series could be obtained. The inequality indices are then measured on those time series. The codes used to produce the simulation results are available in Ref. \cite{code_link}.

\subsubsection{Simulating FBM}
In the FBM, a large number of parallel fibers are connected between two horizontal plates. The top plate with hanging fibers is rigid and the rigidity of the lower plate determines the interaction range of the model. All the fibers are here assumed to have identical elastic constants but have a different failure threshold from each other. Here the failure thresholds of the fibers are randomly assigned from the uniform distribution between (0,1). When a load is applied to the system, the weakest fibers fail irreversibly and the load of the broken fibers is redistributed to the remaining intact fibers. The increased load on the surviving fibers then may cause more failures and this process will go on until the remaining fiber are strong enough to hold the extra load or the whole bundle collapses. The external load is kept constant throughout the redistribution process and it is again increased to initiate the dynamics once the dynamics stops. The number of fibers breaking between two successive load increments is called an avalanche. The redistribution of the load mentioned above depends on the elastic properties of the lower plate. The two cases of load redistribution are equal load sharing (ELS) and Local load sharing (LLS). In ELS, when the lower plate is absolutely rigid, the load of the broken fiber is equally redistributed to all surviving fibers, hence it is the mean field limit. In LLS, when the lower plate deforms under loading, then a higher fraction of the load of broken fiber is carried by the surviving neighbors that are nearer to the broken fiber. The extreme case is the nearest neighbor load sharing. But a long-range (power-law) load sharing has also been studied before \cite{PhysRevE.65.046148}.  Here we redistribute the load of the broken fiber between $R$ surviving neighbors on either side (see e.g., \cite{Subhadeep2017}), when the fibers are arranged in an one-dimensional lattice of length $L$. Clearly, $R=1$ is the nearest neighbor load sharing, and $R\sim L$ is the ELS limit (in fact the ELS limit is reached much earlier \cite{PhysRevResearch.1.033047}). 

Here we study both the mean-field limit (ELS) and the local load sharing limit with finite values of $R$. We mention the respective cases in the results as appropriate.

\subsubsection{Simulating RFM}
The random fuse network model that we consider here is defined on a two-dimensional tilted square lattice of linear size $L$ with periodic boundary conditions along the horizontal direction. A potential difference is applied along the vertical direction between the two opposite sides of the lattice. Each bond $i$ representing an Ohmic resistor with unit conductance carries electrical current $i_i$ until it burns out irreversibly at a threshold value $b_i$ of its current and then, it becomes an insulator. The thresholds values $\{b_i\}$ are selected randomly and independently from a power-law distribution: $p(b) \sim b^{-1}$, bounded between $10^{-\beta}$ and $10^{\beta}$. This is a generic form of distribution with a decaying power-law tail that has been considered widely for modeling fracture of heterogeneous materials with varying degrees of disorder \cite{Moreira2012,Chandreyee2015,Sigmund2015,Subhadeep2017}.

Initially, all the bonds are intact and the applied voltage difference $V$ is raised quasi-statically from zero. The specific geometry of our lattice ensures that every bond initially carries the same amount of current. Therefore, the breaking process initiates when the current through the ``weakest bond" in the system reaches its breaking threshold at $V=V_0=(L-1)\times\min\{b_i\}$. Subsequently, the bond burns and it is irreversibly removed. The new current distribution is then determined by keeping the external voltage difference fixed at $V=V_0$. This may initiate a sequence of bond burning. At this stage, all the bonds carrying currents higher than their respective threshold values are removed simultaneously and again, the current distribution is calculated. Numerically, to determine the current distribution, a set of Kirchhoff's equations are iteratively solved using the conjugate gradient method with an accuracy of $10^{-12}$ \cite{Batrouni}. The technical details of the method and its convergence criteria that we followed are described in Ref.\ \cite{Batrouni}. This way the breaking process continues until a stable state is reached when the current through all the remaining intact bonds is lower than their respective breaking thresholds. This completes an avalanche. The largest ratio between current and threshold, i.e., $\max(i_i/b_i)$ is then calculated for all intact bonds $i$ to determine the next weakest bond to be burnt and removed. This initiates a new avalanche. Accordingly, the external voltage difference is raised. 

In this voltage-controlled setting, the breaking process stops completely when a final crack comprising of burnt bonds wraps around the lattice in a direction transverse to the applied potential difference, i.e., no current flows through the system. The number of bonds that burn between two successive increments of the external voltage determines the avalanche size.

Note that at an early stage of the breaking process, the intact bond having the minimum value of the breaking threshold determines the weakest bond. However, depending on the value of $\beta$, this is not generally true at the later stage. The current concentration around the crack zone competes with the local strength of the breaking thresholds. As a consequence, the breaking events become correlated \cite{Zapperi1997,Moreira2012,Shekhawat,Hansen2003}. Certainly, the choice of $\beta$ is important as discussed in Ref.\ \cite{Moreira2012}. In the limit of weak disorder, i.e., $\beta \to 0$, the width of the distribution is so narrow that the fracture process becomes localized and only a single crack grows in the system, while it becomes percolation-like in the limit of strong disorder, i.e., $\beta \to \infty$. In the later case, one only gets avalanches of size unity and always burns the intact bond with the smallest breaking threshold on the conducting backbone. Importantly, depending on $L$, the breaking dynamics is avalanche dominated in the intermediate range of disorder \cite{Moreira2012,Shekhawat}.

We choose the value of $\beta=$1 and consider system sizes $L=$ 32, 64, 128, 256, and 512 such that the system remains in the transition regime from the weak to the strong disorder where avalanche dynamics can be observed.

\section{Results}
We now move on to describe the behavior of the inequality indices $g$, $h$, and $k$ for the avalanche statistics of the FBM and the RFM. We first present the analytical results for the FBM in the mean-field (ELS) limit and then go on to discuss the local load sharing version and the RFM results.

\subsection{Inequality indices for FBM in mean field limit: Analytical results}

The FBM in the mean-field (ELS) limit is analytically tractable \cite{rmp}. The system size (initial number of fibers) is denoted by $N$, the applied load is $W$, giving the load per fiber $\sigma=W/N$, with $\sigma_c$ denoting the critical value at which catastrophic breakdown happens.  
The fraction of intact fibers at load per fiber $\sigma$, for a threshold distribution uniform in (0,1), is given by
\begin{equation}
U(\sigma)=U^*(\sigma_c)+(\sigma_c-\sigma)^{1/2}.
\label{surviving}
\end{equation}

\begin{figure}
\includegraphics[width=9cm]{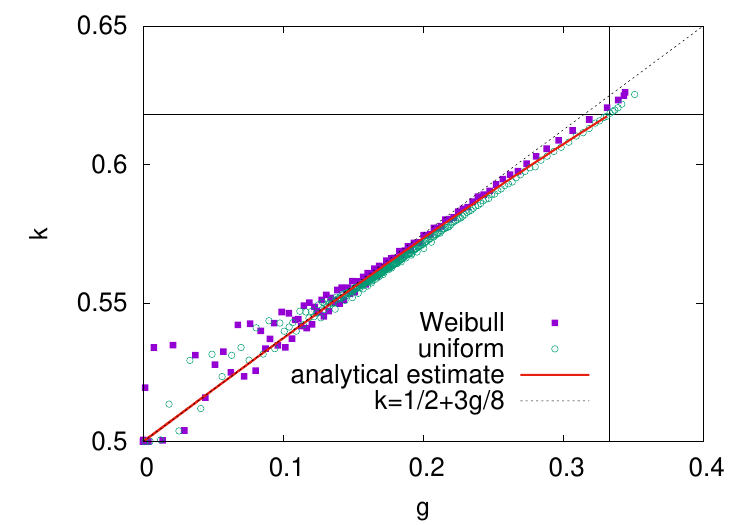}
\caption{The plot of $k(q)$ (obtained numerically from Eq. (\ref{k_exp})) against $g(q)$ (obtained from Eq. (\ref{g_exp})) for the mean field FBM. The numerical simulations of the same are also shown for two threshold distributions -- uniform and Weibull. 
The line $k=\frac{1}{2}+\frac{3}{8}g$ is also plotted for reference, which matches well with the analytical estimate for small values of $g$.
The failure point values of $g$ and $k$, given by $g_f=1/3$ (see Eq. (\ref{limit_g})) and $k_f=(\sqrt{5}-1)/2\approx 0.618$ (see Eq. (\ref{limit_k})), are indicated by a vertical and a horizontal lines, respectively.
 These results indicate that the values of $k$ and $g$ are universal for a broad class of threshold distributions.}
\label{k_vs_g}
\end{figure}

The magnitude of the change in this fraction for a differential increment in the load per fiber value is, therefore,
\begin{equation}
S(\sigma)=\left|\frac{dU(\sigma)}{d\sigma}\right|=\frac{1}{2}(\sigma_c-\sigma)^{-1/2}.
\label{fbm_ava}
\end{equation}
This is a measure of avalanche size at $\sigma$, when the load increment is done by a fixed amount $\Delta=d\sigma$ at every step. 

Note that while Eqs. (\ref{surviving}) \& (\ref{fbm_ava}) are written for uniform threshold distribution in (0,1), they are valid, upto a prefactor, for a broad class of threshold distributions \cite{rmp}. Therefore, all the subsequent results are valid for the same class of threshold distributions. For distributions that are outside this class and can take the failure mode of the bundle from quasi-brittle to an individual fiber breaking dynamics, the inequality estimates for avalanches will also change, as we will see later on.   

\subsubsection{Terminal values of the inequality indices}
The function in Eq. (\ref{fbm_ava}) is monotonically increasing up to the failure point $\sigma=\sigma_c$, hence the Lorenz function at the point of collapse of the system is given by
\begin{equation}
L_f(p)=\frac{\int\limits_0^{p\sigma_c}(\sigma_c-\sigma)^{-1/2}d\sigma}{\int\limits_0^{\sigma_c}(\sigma_c-\sigma)^{-1/2}d\sigma}
\end{equation}

\begin{figure}
    \includegraphics[width=9cm]{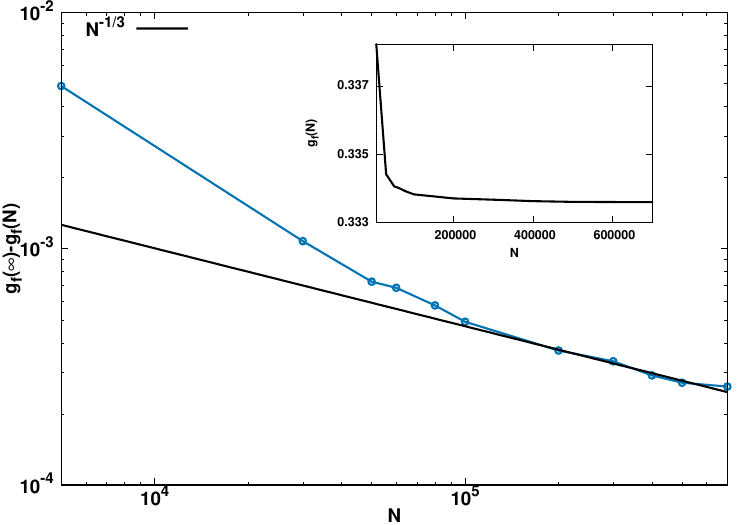}
    \caption{Finite-size scaling of $g$ for system sizes ($N/10^3$ = 5, 30, 50, 60, 80, 100, 200, 300, 400, 500, 700) for the mean-field fiber bundle model. From this figure, we can see that $g_f$ shows the finite size scaling mentioned in Eq. (\ref{fss_gf}) in the limit of large system sizes.}
    \label{fbm_L_g}
\end{figure}

On simplification, this gives
\begin{equation}
L_f(p)=1-\sqrt{1-p},
\label{lorenzt}
\end{equation}
where $\sigma_c$ cancels out, giving the function the robustness it shows against various threshold distributions (see Fig.\ (11) in Ref. \cite{PhysRevE.106.025003}). 
The Gini index ($g$) at the point of collapse is then
\begin{equation}
g_f=1-2\int\limits_0^1L_f(p)dp=\frac{1}{3}.
\label{limit_g}
\end{equation}
The Kolkata index ($k$) at the point of collapse can be found from
\begin{equation}
1-k_f=L_f(k_f)=1-\sqrt{1-k_f},
\end{equation}
which gives
\begin{equation}
k_f=\frac{\sqrt{5}-1}{2}\approx 0.618,
\label{limit_k}
\end{equation}
which is the inverse golden ratio. The other solution is irrelevant. 

It is also known \cite{rmp} that for the fiber bundle model with an equal amount of load increase in each step, the number of fibers broken due to the increase of load between $m\Delta$ to $(m+1)\Delta$ is given by
\begin{equation}
S(m)=\frac{\Delta}{\sqrt{1-4m\Delta/N}},
\label{ava_exp}
\end{equation}
where $m=0,1,2, \dots, N/4\Delta$.
This suggests a monotonic increase in the avalanche size, which is strictly true for equispaced thresholds (given by $i/N$, $i=1, \dots ,N$) but the scaling relations are valid for non-equispaced thresholds as well. To find the Hirsch index ($h$), the avalanches should be first sorted in descending order, which we do with the transformation $m \to \frac{N}{4\Delta}-m$, giving
\begin{equation}
S\left(\frac{N}{4\Delta}-m\right)=\frac{\Delta}{\sqrt{1-\frac{4\Delta}{N}\left(\frac{N}{4\Delta}-m\right)}}=\frac{\Delta}{\sqrt{\frac{4\Delta m}{N}}}.
\end{equation}
Then the value of the Hirsch index at the point of collapse can be found from
\begin{equation}
\frac{\Delta}{\sqrt{\frac{4\Delta h_f}{N}}}=h_f,
\end{equation}
which gives
\begin{equation}
h_f \propto N^{1/3}.
\end{equation}
This scaling is in contrast with the previously conjectured form (see Ref. \cite{k_pre}). However, as we shall shortly see, this is obeyed in accurate numerical simulations of the model.

\subsubsection{Off-critical scaling}
Now, if the loading is done upto $\sigma=q\sigma_c$, where $q\le 1$, then the Lorenz function can be found from
\begin{equation}
L(p,q)=\frac{\int\limits_0^{pq\sigma_c}(\sigma_c-\sigma)^{-1/2}d\sigma}{\int\limits_0^{q\sigma_c}(\sigma_c-\sigma)^{-1/2}d\sigma},
\end{equation}
which gives
\begin{equation}
L(p,q)=\frac{1-\sqrt{1-pq}}{1-\sqrt{1-q}},
\end{equation}
which, in the limit $q=1$, gives back $L_f$ in Eq. (\ref{lorenzt}). It is then possible to calculate the evolution of $g$ and $k$ as the loading is increased, using
\begin{equation}
g(q)=1-2\int\limits_0^1L(p,q)dp,
\end{equation}
which gives on simplification
\begin{equation}
g=1-\frac{4(1-q)\sqrt{1-q}+6q-4}{3q(1-\sqrt{1-q})}.
\label{g_exp}
\end{equation}
Similarly, the equation for $k(q)$ reads
\begin{equation}
k(q)=1-\frac{1-\sqrt{1-k(q)q}}{1-\sqrt{1-q}}.
\label{k_exp}
\end{equation}
For any value of $q$, the value of $k(q)$ can be numerically estimated with arbitrary accuracy from the equation above.
In Fig.\ \ref{k_vs_g}, the values obtained for $k(q)$ from the above equation is plotted against those obtained for $g(q)$ from Eq. (\ref{k_exp}).
A phenomenologically derived  expression $k=\frac{1}{2}+\frac{3}{8}g$ \cite{10.3389/fphy.2022.990278} is also plotted for reference. Furthermore, simulation results for two threshold distributions -- uniform in (0,1) and Weibull (with shape parameter value 1.3) are also shown. For lower values of $g$, for which the phenomenological argument is valid, the two expressions match. For higher values of $g$, the simulations follow the analytical estimate. Given that $g$ and $k$ are almost linearly related, for the subsequent part of this work, we keep the focus on calculating $g$, which can be done in closed form.

\begin{figure}
\includegraphics[width=9cm]{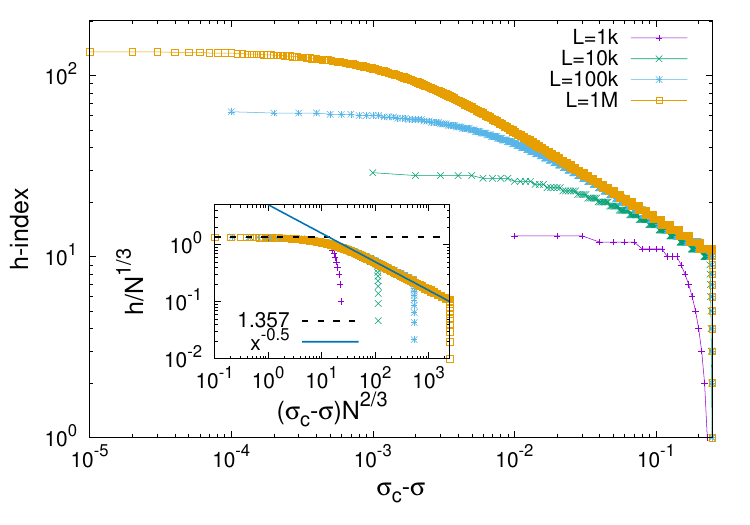}
\caption{The finite size scaling of h-index, following Eqs.\ (22) \& (23) for the mean field FBM. In the inset, the limiting behavior of the scaled function are also indicated -- it saturates to $(5/2)^{1/3}$ with loading step size $\Delta=10$ for small values of the argument and decays as a power-law with exponent $-1/2$ for large values of the argument (see text).}
\label{fss_h}
\end{figure}
\begin{figure*}
\includegraphics[width=17cm]{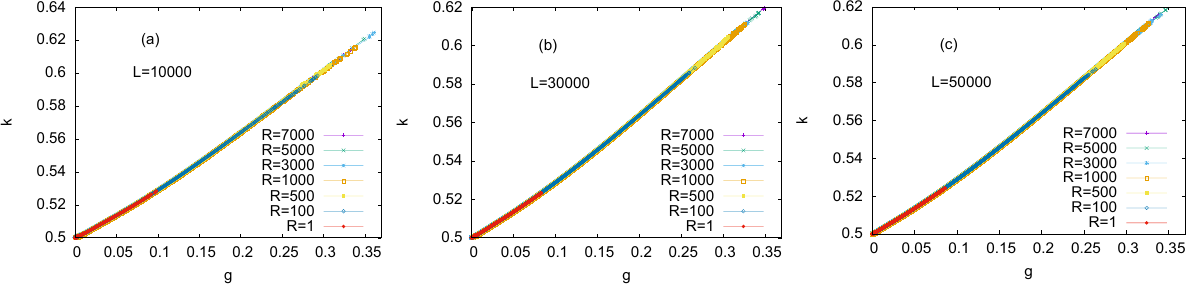}
\caption{The plot of $g(q)$ against $k(q)$ for different interaction ranges $(R = 1, 100, 500, 1000, 3000, 5000, 7000)$ for three different system sizes $(L = 10000, 30000, 50000)$ in LLS fiber bundle model. The plot shows that terminal values of $g$ and $k$ are independent of the system sizes. With increasing the value of $R$, the terminal values of $g$ and $k$ approach to the values obtained analytically in the mean-field limit.}
\label{k_vs_g_LLS}
\end{figure*}
From Eq.(\ref{g_exp}), the $q\to 1$ limit ($\sigma\to\sigma_c$) yields 
\begin{equation}
    g-g_f \propto \sqrt{1-q},
    \label{g_interval}
\end{equation}
meaning that we could write all response functions in terms of the interval $|g-g_f|$ instead of $\sigma-\sigma_c$ with the appropriate changes in the scaling exponent (see Ref.\cite{soumyaditya} for details). An important consequence of the above equation is the finite size scaling of the terminal value of $g$. We could define a correlation length exponent as $\xi \propto |\sigma-\sigma_c|^{-\nu}$ \cite{PhysRevE.102.012113}, which would span the system size at the critical point. We should then have, for a finite size system at its critical point, $|\sigma_c(N)-\sigma_c(N\to\infty)|\propto N^{-1/\nu}$.  Also, at the critical point of the finite system, we should have from Eq. (\ref{g_interval}), 
\begin{equation}
    g_f(N)-g_f(N\to\infty) \propto |\sigma_c(N)-\sigma_c(N\to\infty)|^{1/2}\propto N^{-1/2\nu}.
 \label{fss_gf}
\end{equation}
Given that $\nu=3/2$ \cite{Chandreyee2015,PhysRevE.102.012113}, we should have $|g_f(N)-g_f(N\to\infty)|\propto N^{-1/3}$, which is what we see in Fig.\ \ref{fbm_L_g}.

To find the load and size dependence of $h$, consider again the Eq. (\ref{ava_exp}). Suppose we stop the loading at $m=m_{max}=\frac{qN}{4\Delta}$, where $q\le 1$, as before.
The avalanches arranged in descending order would then read
\begin{equation}
S(m_{max}-m)=\frac{\Delta}{\sqrt{1-\frac{4\Delta}{N}(m_{max}-m)}}=\frac{\Delta}{\sqrt{1-\frac{4\Delta}{N}(\frac{qN}{4\Delta}-m)}}.
\end{equation} 
The Hirsch index can then be found, as before, from
\begin{equation}
\frac{\Delta}{\sqrt{1-\frac{4\Delta}{N}(\frac{qN}{4\Delta}-h)}}=h.
\end{equation}
This can be simplified to
\begin{equation}
h^3+\frac{N(1-q)}{4\Delta}h^2-\frac{N\Delta}{4}=0.
\label{full_h}
\end{equation}
This cubic equation can be solved (see Appendix). But the expression for $h$ is then too complex to figure out the scaling form near the critical point. 
Instead, we can approximate that near the critical point ($q\to 1$), the cubic term in $h$ would be the most dominant, hence responsible for the leading order scaling. In that case, the quadratic term can be replaced by its value exactly at the critical point $h_f^2=\left(\frac{\Delta N}{4}\right)^{2/3}$. 
Then, near the critical point $\sigma_c=1/4$,
\begin{equation}
\frac{4\Delta}{N}h^3\approx \Delta^2-(1-q)\left(\frac{\Delta N}{4}\right)^{2/3}.
\end{equation}
This gives
\begin{equation}
h^3\approx \frac{N\Delta}{4}-\frac{1-q}{4}\frac{N}{\Delta}\left(\frac{\Delta N}{4}\right)^{2/3}.
\end{equation}
On simplification, this gives
\begin{eqnarray}
h^3 &\approx& \frac{\Delta N}{4}-\frac{1-q}{4}\frac{N}{\Delta}\left(\frac{\Delta N}{4}\right)^{2/3} \nonumber \\
    &\approx & N\left[\frac{\Delta}{4}-(\sigma_c-\sigma)N^{2/3}\left(\frac{1}{4\sqrt{\Delta}}\right)^{2/3}\right], 
\end{eqnarray}
where $\frac{1-q}{4}=\sigma_c-\sigma$ (remembering $\sigma=q\sigma_c$). 
This is now in the scaling form
\begin{equation}
h\approx N^{1/3}f\left[(\sigma_c-\sigma)N^{1/\nu}\right],
\label{h_off_scaling}
\end{equation}
with $\nu=3/2$ \cite{Chandreyee2015} being the correlation length exponent.

Fig.\ \ref{fss_h} shows the numerical validation of the scaling form above. The simulation was done for a fixed load increase with $\Delta=10$. So, as $\sigma\to\sigma_c$, the quantity $h/N^{1/3}$ should tend to $(5/2)^{1/3}\approx 1.357$, which it does (as can be seen from the inset of Fig.\ \ref{fss_h}). Also, away from the critical point, the quantity should be proportional to $(\sigma_c - \sigma)^{-1/2}$, which is also seen. The data collapse for the three orders of magnitude system size variation confirms the scaling $h_f\sim N^{1/3}$.

\subsection{Inequality indices in the non-mean field limit of FBM: Simulation results}
\begin{figure*}
   \includegraphics[width=17cm]{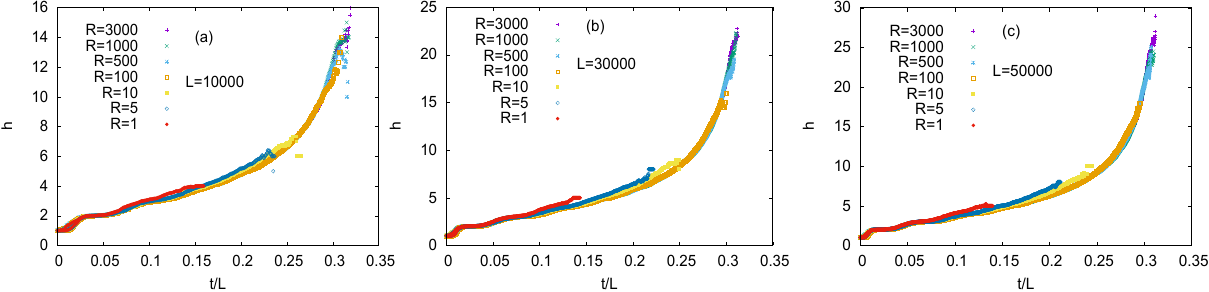}
    \caption{The variation of h-index with re-scaled time $t/L$ for different interaction ranges $(R = 1, 5, 10, 500, 1000, 3000)$ for three different system sizes $(L = 10000, 30000, 50000)$ for LLS fiber bundle model.}
    \label{time_vs_hf}
\end{figure*}

\begin{figure}
    \includegraphics[width=9cm]{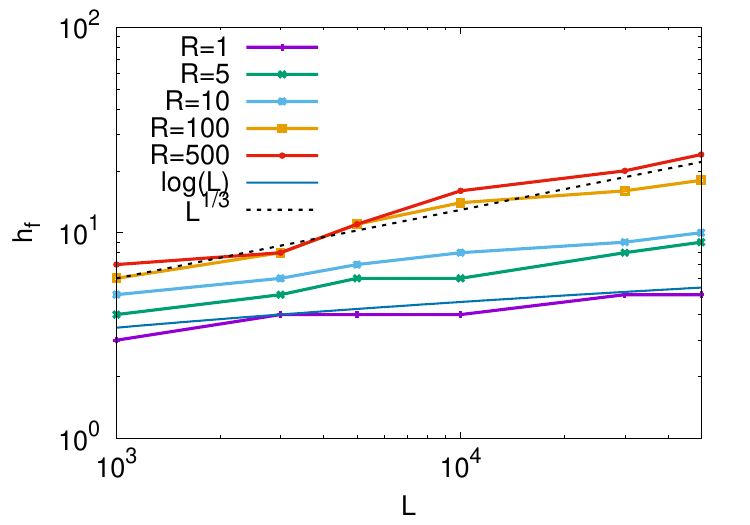}
    \caption{System size $L$ dependence of the terminal values of $h=h_f$ for different interaction ranges $R$ for one dimensional LLS fiber bundle model in a log-log scale. It is seen that for $R=1$ when the load sharing is fully local $h_f$ shows the logarithmic dependence, and when $R=$ 100, and 500, the $h_f$ shows the power law dependence. Here the solid (blue) and dashed (black) lines indicate the $\log(L)$ and $L^{1/3}$ functional form, respectively.}
    \label{L_vs_hf}
\end{figure}

This is the case when a load of a broken fiber is redistributed to the nearest $R$ surviving fibers when they are arranged in a one-dimensional lattice of size $L$.  Here we take the threshold distribution to be uniform between 0 and 1. Here $R = 1$ means, the sharing of the load is fully local. In this case, we take three different system sizes ($L = 10000, 30000, 50000$) and vary the value of $R$, and then calculate the value of $h$, $g$, and $k$. In Fig.\ \ref{k_vs_g_LLS} we plot the $g$ against $k$ and we have observed numerically when the interaction range ($R$) is increased, $g_f$ and $k_f$ values are very close to the analytical results obtained for the mean-field case. Prior to reaching the terminal values, the variation of $g$ vs $k$ almost follows the mean field variation, except that the growth stops earlier (usually there is less number of avalanches for local load sharing). 
In Fig.\ \ref{time_vs_hf} the variation of $h$ is plotted for different values of $R$. Here also, the terminal values are smaller for lower values of $R$, and so are their system size scaling.
Indeed, from the Fig.\ \ref{L_vs_hf} it can be seen that for fully local load sharing ($R=1$), scaling of $h_f$ shows the logarithmic dependence with $L$, but as we increase the interaction range, the scaling of $h_f$ shows the power law dependence of $L^{1/3}$, similar to what we expect in the mean-field limit.

\begin{figure}
    \includegraphics[width=8cm]{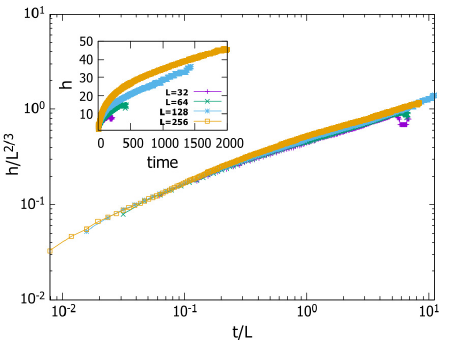}
    \caption{The $h$-index (scaled by $L^{2/3}$) is plotted against the time (scaled by time/L) for different system sizes $(L = 32, 64, 128, 256)$ for the two-dimensional random fuse model with power-law distributed breaking threshold. The inset shows the variation of $h$ with time for the different system sizes.}
    \label{rf_t_h}
\end{figure}
\subsection{Inequality indices for the RFM: Simulation results}
Finally, we compare the above results for a more realistic model of fracture, the random fuse model (RFM) on a tilted square lattice.  In this case, the threshold distribution is a power law with exponent $-1$, as mentioned before. The limits of the distribution are between $10^{-\beta}$ to $10^{\beta}$. We choose the value of $\beta$=1 for the system sizes $L=$ 32, 64, 128, 256 \& 512. We calculated the inequality measures using 500 samples for system sizes up to $L=256$ and 40 samples for $L=512$.
Given that $N=L\times L$, if we expect a similar scaling as in the case of FBM, we should have $h_f\sim L^{2/3}$ (since $N=L\times L$ here). In Fig.\ \ref{rf_t_h}, the function $h/L^{2/3}$ collapses on a single curve, indicating a similar scaling as that in the FBM. 

The values of $g_f$ and $k_f$ are set by the avalanche size distribution exponent. Particularly, if the avalanche size distribution exponent is $\delta$, then it is straightforward to see that when the avalanches are arranged in the ascending order of their sizes, the resulting curve will diverge with an exponent $n=1/(\delta-1)$. For $\beta=1.0$, $\delta\approx 1.99$. Now, it is also possible to show \cite{soumyaditya} that for $n>1$, $g_f=k_f=1$. Therefore, in the finite size scaling of their values, we would have to put $g_f(L\to\infty)\to 1$, resulting in a scaling form as before $|1-g_f(L)|\propto L^{-1/2\nu}$. In Fig.\ \ref{rf_time_g}, we do see a power law decay of $|1-g_f(L)|$ with $L$, but not with the expected exponent value (typically $\nu=4/3$ for RFM \cite{PhysRevE.70.036123}). Nevertheless, the power law scaling is seen, with an exponent $-0.23$.

\section{Discussion and Conclusion}
Near a critical point, the responses of a system show critical divergence. This necessarily means that the responses are highly unequal as a critical point is gradually approached. In the case of disordered solids and their models, these unequal responses are observed in terms of the avalanche time series, which are accumulated local breaking events, experimentally detected as acoustic emissions. It is well known that the statistical properties of such acoustic emissions are useful to characterize the `health' of a system i.e., to estimate the vicinity of an approaching critical (breaking) point.

\begin{figure}
     \includegraphics[width=8cm]{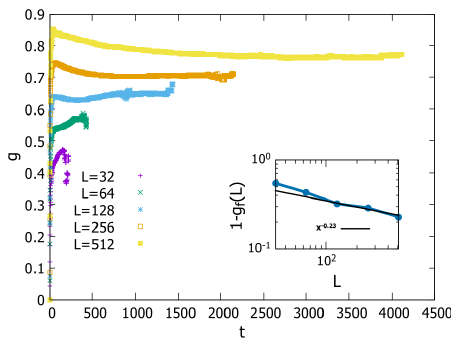}
    \caption{The $g$-index is plotted against time for different system sizes $(L = 32, 64, 128, 256, 512)$ for the random fuse model with power-law distributed breaking threshold. As can be seen from the figure the terminal values of g is changing with the system size. The inset shows the variation of the terminal values of the g-index with time in the log-log scale, exhibiting a power-law scaling.}
    \label{rf_time_g}
\end{figure}

Recently, quantitative measures of the inequality of such avalanches, e.g., the Hirsch index ($h$), the Gini index ($g$), and the Kolkata index ($k$) have been used to forecast the imminent critical point in physical systems \cite{k_pre}. Particularly, it was noted that $g$ and $k$ reach universal terminal values $g_f$ and $k_f$ at the critical point, irrespective of the details of the system (types of disorder present in them). It is therefore very useful to monitor the inequality of the avalanche statistics to estimate the distance from an approaching critical point \cite{PhysRevE.106.025003}.  It is more useful to monitor the inequality indices than the other universal quantities e.g., the avalanche size distribution critical exponent because such exponent shows universal properties only in the asymptotic limit of the large avalanche sizes, where the failure is already extremely imminent. Indeed, extensive numerical simulations and machine learning studies have indicated the same \cite{PhysRevE.106.025003}. It shows that for experimental measurements, inequalities of the acoustic emissions, and not just their sizes, could play an important role for the indication of the approaching failure point.

Here we have shown analytically, in the mean-field FBM, the scaling behavior of the critical indices ($h$, $k$, and $g$). Particularly, for the terminal values of the inequality indices we have shown that $h_f\sim N^{1/3}$, where $N$ is the system size, $g_f=1/3$ and $k_f=(\sqrt{5}-1)/2$. We have also shown the off-critical Widom-Stauffer like scaling for $h$ (see Eq. (\ref{h_off_scaling})). These are also verified numerically (see Figs \ref{fss_h} \& \ref{L_vs_hf}). Some of these scaling relations differ from what was conjectured using only numerical simulations before \cite{k_pre}. 


We then study the more realistic local load sharing non-mean field version of the FBM and also the RFM. In LLS-FBM, we see that the scaling of $h_f$ shifts from a logarithmic dependence for very localized load sharing, to the power-law dependence mentioned above, as the load sharing range $R$ is increased. However, the power-law scaling with system size is observed for the RFM (see Fig \ref{rf_t_h}). We also show the finite size effect in $g_f$ for the FBM and the RFM (see Figs \ref{fbm_L_g} \& \ref{rf_time_g} ).

In conclusion, the inequalities of the avalanche sizes in models of fracture show universal scaling properties near the failure point. Here we show analytically (for mean field FBM) and numerically (for non-mean field FBM and RFM) the scaling properties of the inequality indices. These results show that monitoring the inequality indices in stressed disordered solids can indicate approaching a critical point, as was conjectured from numerical and machine learning analysis before.

\section*{Acknowledgement}
SK is supported by the research grant `ORLA$\_$BIRD2020$\_$01' of the University of Padova. BKC is grateful to the Indian
National Science Academy for their Senior Scientist Research Grant. The simulations of FBM were done using HPCC Surya in SRM University - AP.

\section*{Appendix: The exact solution for $h$}
As mentioned before, a full solution of Eq. (\ref{full_h}) is possible \cite{cubic}. 
The equation can be compared with a general cubic form $ah^3+bh^2+ch+d=0$, with $a=1$, $b=\frac{N(1-q)}{4\Delta}$, $c=0$ and $d=-\frac{N\Delta}{4}$. Let us now define two quantities $D_0=b^2-3ac=\frac{N^2(1-q)^2}{16\Delta^2}$ and $D_1=2b^3-9abc+27a^2d=\frac{N^3(1-q)^3}{32\Delta^3}-\frac{27N\Delta}{4}$.
Then consider $C=\sqrt[3]{\frac{D_1\pm \sqrt{D_1^2-4D_0^3}}{2}}$. 

Finally, the three roots are 
\begin{equation}
    h=-\frac{1}{3}\left(\frac{N(1-q)}{4\Delta}+\zeta^mC+\frac{D_0}{\zeta^mC} \right),
\end{equation}
with $m\in \{1,2,3\}$ and $\zeta=\frac{-1+\sqrt{-3}}{2}$. Of course, only the real root will be physically meaningful.


\begin{thebibliography}{37}%
\makeatletter
\providecommand \@ifxundefined [1]{%
 \@ifx{#1\undefined}
}%
\providecommand \@ifnum [1]{%
 \ifnum #1\expandafter \@firstoftwo
 \else \expandafter \@secondoftwo
 \fi
}%
\providecommand \@ifx [1]{%
 \ifx #1\expandafter \@firstoftwo
 \else \expandafter \@secondoftwo
 \fi
}%
\providecommand \natexlab [1]{#1}%
\providecommand \enquote  [1]{``#1''}%
\providecommand \bibnamefont  [1]{#1}%
\providecommand \bibfnamefont [1]{#1}%
\providecommand \citenamefont [1]{#1}%
\providecommand \href@noop [0]{\@secondoftwo}%
\providecommand \href [0]{\begingroup \@sanitize@url \@href}%
\providecommand \@href[1]{\@@startlink{#1}\@@href}%
\providecommand \@@href[1]{\endgroup#1\@@endlink}%
\providecommand \@sanitize@url [0]{\catcode `\\12\catcode `\$12\catcode
  `\&12\catcode `\#12\catcode `\^12\catcode `\_12\catcode `\%12\relax}%
\providecommand \@@startlink[1]{}%
\providecommand \@@endlink[0]{}%
\providecommand \url  [0]{\begingroup\@sanitize@url \@url }%
\providecommand \@url [1]{\endgroup\@href {#1}{\urlprefix }}%
\providecommand \urlprefix  [0]{URL }%
\providecommand \Eprint [0]{\href }%
\providecommand \doibase [0]{http://dx.doi.org/}%
\providecommand \selectlanguage [0]{\@gobble}%
\providecommand \bibinfo  [0]{\@secondoftwo}%
\providecommand \bibfield  [0]{\@secondoftwo}%
\providecommand \translation [1]{[#1]}%
\providecommand \BibitemOpen [0]{}%
\providecommand \bibitemStop [0]{}%
\providecommand \bibitemNoStop [0]{.\EOS\space}%
\providecommand \EOS [0]{\spacefactor3000\relax}%
\providecommand \BibitemShut  [1]{\csname bibitem#1\endcsname}%
\let\auto@bib@innerbib\@empty
\bibitem [{\citenamefont {Biswas}\ \emph {et~al.}(2015)\citenamefont {Biswas},
  \citenamefont {Ray},\ and\ \citenamefont {Chakrabarti}}]{wiley_book}%
  \BibitemOpen
  \bibfield  {author} {\bibinfo {author} {\bibfnamefont {S.}~\bibnamefont
  {Biswas}}, \bibinfo {author} {\bibfnamefont {P.}~\bibnamefont {Ray}}, \ and\
  \bibinfo {author} {\bibfnamefont {B.~K.}\ \bibnamefont {Chakrabarti}},\
  }\href@noop {} {\emph {\bibinfo {title} {Statistical physics of fracture,
  breakdown and earthquake}}}\ (\bibinfo  {publisher} {Wiley},\ \bibinfo {year}
  {2015})\BibitemShut {NoStop}%
\bibitem [{\citenamefont {Bonamy}\ and\ \citenamefont
  {Bouchaud}(2011)}]{BONAMY20111}%
  \BibitemOpen
  \bibfield  {author} {\bibinfo {author} {\bibfnamefont {D.}~\bibnamefont
  {Bonamy}}\ and\ \bibinfo {author} {\bibfnamefont {E.}~\bibnamefont
  {Bouchaud}},\ }\bibfield  {title} {\enquote {\bibinfo {title} {Failure of
  heterogeneous materials: A dynamic phase transition?}}\ }\href {\doibase
  https://doi.org/10.1016/j.physrep.2010.07.006} {\bibfield  {journal}
  {\bibinfo  {journal} {Physics Reports}\ }\textbf {\bibinfo {volume} {498}},\
  \bibinfo {pages} {1--44} (\bibinfo {year} {2011})}\BibitemShut {NoStop}%
\bibitem [{\citenamefont {Sethna}\ \emph {et~al.}(2001)\citenamefont {Sethna},
  \citenamefont {Dahmen},\ and\ \citenamefont {Myers}}]{crackling}%
  \BibitemOpen
  \bibfield  {author} {\bibinfo {author} {\bibfnamefont {James~P.}\
  \bibnamefont {Sethna}}, \bibinfo {author} {\bibfnamefont {Karin~A.}\
  \bibnamefont {Dahmen}}, \ and\ \bibinfo {author} {\bibfnamefont
  {Christopher~R.}\ \bibnamefont {Myers}},\ }\bibfield  {title} {\enquote
  {\bibinfo {title} {Crackling noise},}\ }\href {\doibase
  https://doi.org/10.1038/35065675} {\bibfield  {journal} {\bibinfo  {journal}
  {Nature}\ }\textbf {\bibinfo {volume} {410}},\ \bibinfo {pages} {242--250}
  (\bibinfo {year} {2001})}\BibitemShut {NoStop}%
\bibitem [{\citenamefont {Gutenberg}\ and\ \citenamefont
  {F.~Richter}(1949)}]{earthquake}%
  \BibitemOpen
  \bibfield  {author} {\bibinfo {author} {\bibfnamefont {B.}~\bibnamefont
  {Gutenberg}}\ and\ \bibinfo {author} {\bibfnamefont {C.}~\bibnamefont
  {F.~Richter}},\ }\bibfield  {title} {\enquote {\bibinfo {title} {Seismicity
  of the earth and associated phenomena},}\ }\href
  {https://archive.org/details/seismicityofthee009299mbp/page/n5/mode/1up}
  {\bibfield  {journal} {\bibinfo  {journal} {Princeton University Press}\ ,\
  \bibinfo {pages} {295}} (\bibinfo {year} {1949})}\BibitemShut {NoStop}%
\bibitem [{\citenamefont {Chatterjee}\ \emph {et~al.}(2015)\citenamefont
  {Chatterjee}, \citenamefont {Ghosh}, \citenamefont {ichi Inoue},\ and\
  \citenamefont {Chakrabarti}}]{Chatterjee_2015}%
  \BibitemOpen
  \bibfield  {author} {\bibinfo {author} {\bibfnamefont {Arnab}\ \bibnamefont
  {Chatterjee}}, \bibinfo {author} {\bibfnamefont {Asim}\ \bibnamefont
  {Ghosh}}, \bibinfo {author} {\bibfnamefont {Jun}\ \bibnamefont {ichi Inoue}},
  \ and\ \bibinfo {author} {\bibfnamefont {Bikas~K}\ \bibnamefont
  {Chakrabarti}},\ }\bibfield  {title} {\enquote {\bibinfo {title} {Social
  inequality: from data to statistical physics modeling},}\ }\href {\doibase
  10.1088/1742-6596/638/1/012014} {\bibfield  {journal} {\bibinfo  {journal}
  {Journal of Physics: Conference Series}\ }\textbf {\bibinfo {volume} {638}},\
  \bibinfo {pages} {012014} (\bibinfo {year} {2015})}\BibitemShut {NoStop}%
\bibitem [{\citenamefont {Chakrabarti}\ \emph {et~al.}(2013)\citenamefont
  {Chakrabarti}, \citenamefont {Chakraborti}, \citenamefont {Chakravarty},\
  and\ \citenamefont {Chatterjee}}]{chakrabarti2013econophysics}%
  \BibitemOpen
  \bibfield  {author} {\bibinfo {author} {\bibfnamefont {B.K.}\ \bibnamefont
  {Chakrabarti}}, \bibinfo {author} {\bibfnamefont {A.}~\bibnamefont
  {Chakraborti}}, \bibinfo {author} {\bibfnamefont {S.R.}\ \bibnamefont
  {Chakravarty}}, \ and\ \bibinfo {author} {\bibfnamefont {A.}~\bibnamefont
  {Chatterjee}},\ }\href {https://books.google.co.in/books?id=LWAgAwAAQBAJ}
  {\emph {\bibinfo {title} {Econophysics of Income and Wealth Distributions}}}\
  (\bibinfo  {publisher} {Cambridge University Press},\ \bibinfo {year}
  {2013})\BibitemShut {NoStop}%
\bibitem [{\citenamefont {Bak}\ \emph {et~al.}(1987)\citenamefont {Bak},
  \citenamefont {Tang},\ and\ \citenamefont {Wiesenfeld}}]{PhysRevLett.59.381}%
  \BibitemOpen
  \bibfield  {author} {\bibinfo {author} {\bibfnamefont {Per}\ \bibnamefont
  {Bak}}, \bibinfo {author} {\bibfnamefont {Chao}\ \bibnamefont {Tang}}, \ and\
  \bibinfo {author} {\bibfnamefont {Kurt}\ \bibnamefont {Wiesenfeld}},\
  }\bibfield  {title} {\enquote {\bibinfo {title} {Self-organized criticality:
  An explanation of the 1/f noise},}\ }\href {\doibase
  10.1103/PhysRevLett.59.381} {\bibfield  {journal} {\bibinfo  {journal} {Phys.
  Rev. Lett.}\ }\textbf {\bibinfo {volume} {59}},\ \bibinfo {pages} {381--384}
  (\bibinfo {year} {1987})}\BibitemShut {NoStop}%
\bibitem [{\citenamefont {Pareto}(1896-1965)}]{Pareto}%
  \BibitemOpen
  \bibfield  {author} {\bibinfo {author} {\bibfnamefont {V.}~\bibnamefont
  {Pareto}},\ }\bibfield  {title} {\enquote {\bibinfo {title} {Cours d'
  economie politique},}\ }\href
  {https://www.britannica.com/biography/Vilfredo-Pareto} {\bibfield  {journal}
  {\bibinfo  {journal} {Reprinted as a volume of Oeuvres Compl‘etes}\ }
  (\bibinfo {year} {1896-1965})}\BibitemShut {NoStop}%
\bibitem [{\citenamefont {Bourguignon}(2015)}]{inequality}%
  \BibitemOpen
  \bibfield  {author} {\bibinfo {author} {\bibfnamefont {F.}~\bibnamefont
  {Bourguignon}},\ }\bibfield  {title} {\enquote {\bibinfo {title}
  {Globalization of inequality},}\ }\href@noop {} {\bibfield  {journal}
  {\bibinfo  {journal} {Princeton University Press}\ } (\bibinfo {year}
  {2015})}\BibitemShut {NoStop}%
\bibitem [{\citenamefont {Biswas}\ and\ \citenamefont
  {Chakrabarti}(2021)}]{k_pre}%
  \BibitemOpen
  \bibfield  {author} {\bibinfo {author} {\bibfnamefont {Soumyajyoti}\
  \bibnamefont {Biswas}}\ and\ \bibinfo {author} {\bibfnamefont {Bikas~K.}\
  \bibnamefont {Chakrabarti}},\ }\bibfield  {title} {\enquote {\bibinfo {title}
  {Social inequality analysis of fiber bundle model statistics and prediction
  of materials failure},}\ }\href {\doibase 10.1103/PhysRevE.104.044308}
  {\bibfield  {journal} {\bibinfo  {journal} {Phys. Rev. E}\ }\textbf {\bibinfo
  {volume} {104}},\ \bibinfo {pages} {044308} (\bibinfo {year}
  {2021})}\BibitemShut {NoStop}%
\bibitem [{\citenamefont {Diksha}\ and\ \citenamefont
  {Biswas}(2022)}]{PhysRevE.106.025003}%
  \BibitemOpen
  \bibfield  {author} {\bibinfo {author} {\bibnamefont {Diksha}}\ and\ \bibinfo
  {author} {\bibfnamefont {Soumyajyoti}\ \bibnamefont {Biswas}},\ }\bibfield
  {title} {\enquote {\bibinfo {title} {Prediction of imminent failure using
  supervised learning in a fiber bundle model},}\ }\href {\doibase
  10.1103/PhysRevE.106.025003} {\bibfield  {journal} {\bibinfo  {journal}
  {Phys. Rev. E}\ }\textbf {\bibinfo {volume} {106}},\ \bibinfo {pages}
  {025003} (\bibinfo {year} {2022})}\BibitemShut {NoStop}%
  \bibitem{code_link}
  https://github.com/soumya-84/FBM\_RFM
\bibitem [{\citenamefont {Manna}\ \emph {et~al.}(2022)\citenamefont {Manna},
  \citenamefont {Biswas},\ and\ \citenamefont {Chakrabarti}}]{soc_fbm}%
  \BibitemOpen
  \bibfield  {author} {\bibinfo {author} {\bibfnamefont {S.S.}\ \bibnamefont
  {Manna}}, \bibinfo {author} {\bibfnamefont {Soumyajyoti}\ \bibnamefont
  {Biswas}}, \ and\ \bibinfo {author} {\bibfnamefont {Bikas~K.}\ \bibnamefont
  {Chakrabarti}},\ }\bibfield  {title} {\enquote {\bibinfo {title} {Near
  universal values of social inequality indices in self-organized critical
  models},}\ }\href {\doibase https://doi.org/10.1016/j.physa.2022.127121}
  {\bibfield  {journal} {\bibinfo  {journal} {Physica A: Statistical Mechanics
  and its Applications}\ }\textbf {\bibinfo {volume} {596}},\ \bibinfo {pages}
  {127121} (\bibinfo {year} {2022})}\BibitemShut {NoStop}%
\bibitem [{\citenamefont {Ghosh}\ \emph {et~al.}(2022)\citenamefont {Ghosh},
  \citenamefont {Biswas},\ and\ \citenamefont
  {Chakrabarti}}]{10.3389/fphy.2022.990278}%
  \BibitemOpen
  \bibfield  {author} {\bibinfo {author} {\bibfnamefont {Asim}\ \bibnamefont
  {Ghosh}}, \bibinfo {author} {\bibfnamefont {Soumyajyoti}\ \bibnamefont
  {Biswas}}, \ and\ \bibinfo {author} {\bibfnamefont {Bikas~K.}\ \bibnamefont
  {Chakrabarti}},\ }\bibfield  {title} {\enquote {\bibinfo {title} {Success of
  social inequality measures in predicting critical or failure points in some
  models of physical systems},}\ }\href {\doibase 10.3389/fphy.2022.990278}
  {\bibfield  {journal} {\bibinfo  {journal} {Frontiers in Physics}\ }\textbf
  {\bibinfo {volume} {10}} (\bibinfo {year} {2022}),\
  10.3389/fphy.2022.990278}\BibitemShut {NoStop}%
\bibitem [{\citenamefont {Dalton}(1920)}]{gini}%
  \BibitemOpen
  \bibfield  {author} {\bibinfo {author} {\bibfnamefont {Hugh}\ \bibnamefont
  {Dalton}},\ }\bibfield  {title} {\enquote {\bibinfo {title} {The measurement
  of the inequality of incomes},}\ }\href {http://www.jstor.org/stable/2223525}
  {\bibfield  {journal} {\bibinfo  {journal} {The Economic Journal}\ }\textbf
  {\bibinfo {volume} {30}},\ \bibinfo {pages} {348--361} (\bibinfo {year}
  {1920})}\BibitemShut {NoStop}%
\bibitem [{\citenamefont {Hirsch}(2005)}]{h-index}%
  \BibitemOpen
  \bibfield  {author} {\bibinfo {author} {\bibfnamefont {J.~E.}\ \bibnamefont
  {Hirsch}},\ }\bibfield  {title} {\enquote {\bibinfo {title} {An index to
  quantify an individual's scientific research output},}\ }\href {\doibase
  10.1073/pnas.0507655102} {\bibfield  {journal} {\bibinfo  {journal}
  {Proceedings of the National Academy of Sciences}\ }\textbf {\bibinfo
  {volume} {102}},\ \bibinfo {pages} {16569--16572} (\bibinfo {year}
  {2005})}\BibitemShut {NoStop}%
\bibitem [{\citenamefont {Ghosh}\ \emph {et~al.}(2014)\citenamefont {Ghosh},
  \citenamefont {Chattopadhyay},\ and\ \citenamefont {Chakrabarti}}]{kolkata}%
  \BibitemOpen
  \bibfield  {author} {\bibinfo {author} {\bibfnamefont {Asim}\ \bibnamefont
  {Ghosh}}, \bibinfo {author} {\bibfnamefont {Nachiketa}\ \bibnamefont
  {Chattopadhyay}}, \ and\ \bibinfo {author} {\bibfnamefont {Bikas~K.}\
  \bibnamefont {Chakrabarti}},\ }\bibfield  {title} {\enquote {\bibinfo {title}
  {Inequality in societies, academic institutions and science journals: Gini
  and k-indices},}\ }\href {\doibase
  https://doi.org/10.1016/j.physa.2014.05.026} {\bibfield  {journal} {\bibinfo
  {journal} {Physica A: Statistical Mechanics and its Applications}\ }\textbf
  {\bibinfo {volume} {410}},\ \bibinfo {pages} {30--34} (\bibinfo {year}
  {2014})}\BibitemShut {NoStop}%
\bibitem [{\citenamefont {T.~Pierce}(1926)}]{pierce}%
  \BibitemOpen
  \bibfield  {author} {\bibinfo {author} {\bibfnamefont {F.}~\bibnamefont
  {T.~Pierce}},\ }\bibfield  {title} {\enquote {\bibinfo {title}
  {32—x.—tensile tests for cotton yarns v.—“the weakest link”
  theorems on the strength of long and of composite specimens},}\ }\href
  {\doibase 10.1080/19447027.1926.10599953} {\bibfield  {journal} {\bibinfo
  {journal} {Journal of the Textile Institute Transactions}\ }\textbf {\bibinfo
  {volume} {17}},\ \bibinfo {pages} {T355--T368} (\bibinfo {year}
  {1926})}\BibitemShut {NoStop}%
\bibitem [{\citenamefont {Pradhan}\ \emph {et~al.}(2010)\citenamefont
  {Pradhan}, \citenamefont {Hansen},\ and\ \citenamefont {Chakrabarti}}]{rmp}%
  \BibitemOpen
  \bibfield  {author} {\bibinfo {author} {\bibfnamefont {Srutarshi}\
  \bibnamefont {Pradhan}}, \bibinfo {author} {\bibfnamefont {Alex}\
  \bibnamefont {Hansen}}, \ and\ \bibinfo {author} {\bibfnamefont {Bikas~K.}\
  \bibnamefont {Chakrabarti}},\ }\bibfield  {title} {\enquote {\bibinfo {title}
  {Failure processes in elastic fiber bundles},}\ }\href {\doibase
  10.1103/RevModPhys.82.499} {\bibfield  {journal} {\bibinfo  {journal} {Rev.
  Mod. Phys.}\ }\textbf {\bibinfo {volume} {82}},\ \bibinfo {pages} {499--555}
  (\bibinfo {year} {2010})}\BibitemShut {NoStop}%
\bibitem [{\citenamefont {{de Arcangelis, L.}}\ \emph
  {et~al.}(1985)\citenamefont {{de Arcangelis, L.}}, \citenamefont {{Redner,
  S.}},\ and\ \citenamefont {{Herrmann, H. J.}}}]{Herrmann}%
  \BibitemOpen
  \bibfield  {author} {\bibinfo {author} {\bibnamefont {{de Arcangelis, L.}}},
  \bibinfo {author} {\bibnamefont {{Redner, S.}}}, \ and\ \bibinfo {author}
  {\bibnamefont {{Herrmann, H. J.}}},\ }\bibfield  {title} {\enquote {\bibinfo
  {title} {A random fuse model for breaking processes},}\ }\href {\doibase
  10.1051/jphyslet:019850046013058500} {\bibfield  {journal} {\bibinfo
  {journal} {J. Physique Lett.}\ }\textbf {\bibinfo {volume} {46}},\ \bibinfo
  {pages} {585--590} (\bibinfo {year} {1985})}\BibitemShut {NoStop}%
\bibitem [{\citenamefont {Kahng}\ \emph {et~al.}(1988)\citenamefont {Kahng},
  \citenamefont {Batrouni}, \citenamefont {Redner}, \citenamefont
  {de~Arcangelis},\ and\ \citenamefont {Herrmann}}]{Kahng}%
  \BibitemOpen
  \bibfield  {author} {\bibinfo {author} {\bibfnamefont {B.}~\bibnamefont
  {Kahng}}, \bibinfo {author} {\bibfnamefont {G.~G.}\ \bibnamefont {Batrouni}},
  \bibinfo {author} {\bibfnamefont {S.}~\bibnamefont {Redner}}, \bibinfo
  {author} {\bibfnamefont {L.}~\bibnamefont {de~Arcangelis}}, \ and\ \bibinfo
  {author} {\bibfnamefont {H.~J.}\ \bibnamefont {Herrmann}},\ }\bibfield
  {title} {\enquote {\bibinfo {title} {Electrical breakdown in a fuse network
  with random, continuously distributed breaking strengths},}\ }\href {\doibase
  10.1103/PhysRevB.37.7625} {\bibfield  {journal} {\bibinfo  {journal} {Phys.
  Rev. B}\ }\textbf {\bibinfo {volume} {37}},\ \bibinfo {pages} {7625--7637}
  (\bibinfo {year} {1988})}\BibitemShut {NoStop}%
\bibitem [{\citenamefont {Zapperi}\ \emph {et~al.}(1997)\citenamefont
  {Zapperi}, \citenamefont {Ray}, \citenamefont {Stanley},\ and\ \citenamefont
  {Vespignani}}]{Zapperi1997}%
  \BibitemOpen
  \bibfield  {author} {\bibinfo {author} {\bibfnamefont {Stefano}\ \bibnamefont
  {Zapperi}}, \bibinfo {author} {\bibfnamefont {Purusattam}\ \bibnamefont
  {Ray}}, \bibinfo {author} {\bibfnamefont {H.~Eugene}\ \bibnamefont
  {Stanley}}, \ and\ \bibinfo {author} {\bibfnamefont {Alessandro}\
  \bibnamefont {Vespignani}},\ }\bibfield  {title} {\enquote {\bibinfo {title}
  {First-order transition in the breakdown of disordered media},}\ }\href
  {\doibase 10.1103/PhysRevLett.78.1408} {\bibfield  {journal} {\bibinfo
  {journal} {Phys. Rev. Lett.}\ }\textbf {\bibinfo {volume} {78}},\ \bibinfo
  {pages} {1408--1411} (\bibinfo {year} {1997})}\BibitemShut {NoStop}%
\bibitem [{\citenamefont {Alava}\ \emph {et~al.}(2006)\citenamefont {Alava},
  \citenamefont {Nukala},\ and\ \citenamefont
  {Zapperi}}]{doi:10.1080/00018730300741518}%
  \BibitemOpen
  \bibfield  {author} {\bibinfo {author} {\bibfnamefont {Mikko~J.}\
  \bibnamefont {Alava}}, \bibinfo {author} {\bibfnamefont {Phani K. V.~V.}\
  \bibnamefont {Nukala}}, \ and\ \bibinfo {author} {\bibfnamefont {Stefano}\
  \bibnamefont {Zapperi}},\ }\bibfield  {title} {\enquote {\bibinfo {title}
  {Statistical models of fracture},}\ }\href {\doibase
  10.1080/00018730300741518} {\bibfield  {journal} {\bibinfo  {journal}
  {Advances in Physics}\ }\textbf {\bibinfo {volume} {55}},\ \bibinfo {pages}
  {349--476} (\bibinfo {year} {2006})},\ \Eprint
  {http://arxiv.org/abs/https://doi.org/10.1080/00018730300741518}
  {https://doi.org/10.1080/00018730300741518} \BibitemShut {NoStop}%
\bibitem [{\citenamefont {Hansen}\ \emph {et~al.}(2015)\citenamefont {Hansen},
  \citenamefont {Hemmer},\ and\ \citenamefont {Pradhan}}]{alex_book}%
  \BibitemOpen
  \bibfield  {author} {\bibinfo {author} {\bibfnamefont {Alex}\ \bibnamefont
  {Hansen}}, \bibinfo {author} {\bibfnamefont {Per~Christian}\ \bibnamefont
  {Hemmer}}, \ and\ \bibinfo {author} {\bibfnamefont {Srutarshi}\ \bibnamefont
  {Pradhan}},\ }\href@noop {} {\emph {\bibinfo {title} {The Fiber Bundle Model:
  Modeling Failure in Materials}}}\ (\bibinfo  {publisher} {Wiley},\ \bibinfo
  {year} {2015})\BibitemShut {NoStop}%

\bibitem{metal}
M. Lebyodkin, T. Lebedkina, {\it On the Role of Hazard and Particle Failure Statistics on the Variation of Fracture Parameters of Ductile-Brittle Composites},
Metals {\bf 9}, 633 (2019).

\bibitem [{\citenamefont {Pradhan}\ \emph {et~al.}(2005)\citenamefont
  {Pradhan}, \citenamefont {Hansen},\ and\ \citenamefont
  {Hemmer}}]{PhysRevLett.95.125501}%
  \BibitemOpen
  \bibfield  {author} {\bibinfo {author} {\bibfnamefont {Srutarshi}\
  \bibnamefont {Pradhan}}, \bibinfo {author} {\bibfnamefont {Alex}\
  \bibnamefont {Hansen}}, \ and\ \bibinfo {author} {\bibfnamefont {Per~C.}\
  \bibnamefont {Hemmer}},\ }\bibfield  {title} {\enquote {\bibinfo {title}
  {Crossover behavior in burst avalanches: Signature of imminent failure},}\
  }\href {\doibase 10.1103/PhysRevLett.95.125501} {\bibfield  {journal}
  {\bibinfo  {journal} {Phys. Rev. Lett.}\ }\textbf {\bibinfo {volume} {95}},\
  \bibinfo {pages} {125501} (\bibinfo {year} {2005})}\BibitemShut {NoStop}%
\bibitem [{\citenamefont {Pradhan}\ \emph {et~al.}(2019)\citenamefont
  {Pradhan}, \citenamefont {Kjellstadli},\ and\ \citenamefont
  {Hansen}}]{10.3389/fphy.2019.00106}%
  \BibitemOpen
  \bibfield  {author} {\bibinfo {author} {\bibfnamefont {Srutarshi}\
  \bibnamefont {Pradhan}}, \bibinfo {author} {\bibfnamefont {Jonas~T.}\
  \bibnamefont {Kjellstadli}}, \ and\ \bibinfo {author} {\bibfnamefont {Alex}\
  \bibnamefont {Hansen}},\ }\bibfield  {title} {\enquote {\bibinfo {title}
  {Variation of elastic energy shows reliable signal of upcoming catastrophic
  failure},}\ }\href {\doibase 10.3389/fphy.2019.00106} {\bibfield  {journal}
  {\bibinfo  {journal} {Frontiers in Physics}\ }\textbf {\bibinfo {volume} {7}}
  (\bibinfo {year} {2019}),\ 10.3389/fphy.2019.00106}\BibitemShut {NoStop}%
\bibitem [{\citenamefont {Hidalgo}\ \emph {et~al.}(2002)\citenamefont
  {Hidalgo}, \citenamefont {Moreno}, \citenamefont {Kun},\ and\ \citenamefont
  {Herrmann}}]{PhysRevE.65.046148}%
  \BibitemOpen
  \bibfield  {author} {\bibinfo {author} {\bibfnamefont {Raul~Cruz}\
  \bibnamefont {Hidalgo}}, \bibinfo {author} {\bibfnamefont {Yamir}\
  \bibnamefont {Moreno}}, \bibinfo {author} {\bibfnamefont {Ferenc}\
  \bibnamefont {Kun}}, \ and\ \bibinfo {author} {\bibfnamefont {Hans~J.}\
  \bibnamefont {Herrmann}},\ }\bibfield  {title} {\enquote {\bibinfo {title}
  {Fracture model with variable range of interaction},}\ }\href {\doibase
  10.1103/PhysRevE.65.046148} {\bibfield  {journal} {\bibinfo  {journal} {Phys.
  Rev. E}\ }\textbf {\bibinfo {volume} {65}},\ \bibinfo {pages} {046148}
  (\bibinfo {year} {2002})}\BibitemShut {NoStop}%
\bibitem [{\citenamefont {Roy}\ \emph {et~al.}(2017)\citenamefont {Roy},
  \citenamefont {Biswas},\ and\ \citenamefont {Ray}}]{Subhadeep2017}%
  \BibitemOpen
  \bibfield  {author} {\bibinfo {author} {\bibfnamefont {Subhadeep}\
  \bibnamefont {Roy}}, \bibinfo {author} {\bibfnamefont {Soumyajyoti}\
  \bibnamefont {Biswas}}, \ and\ \bibinfo {author} {\bibfnamefont {Purusattam}\
  \bibnamefont {Ray}},\ }\bibfield  {title} {\enquote {\bibinfo {title} {Modes
  of failure in disordered solids},}\ }\href {\doibase
  10.1103/PhysRevE.96.063003} {\bibfield  {journal} {\bibinfo  {journal} {Phys.
  Rev. E}\ }\textbf {\bibinfo {volume} {96}},\ \bibinfo {pages} {063003}
  (\bibinfo {year} {2017})}\BibitemShut {NoStop}%
\bibitem [{\citenamefont {Roy}\ \emph {et~al.}(2019)\citenamefont {Roy},
  \citenamefont {Biswas},\ and\ \citenamefont
  {Ray}}]{PhysRevResearch.1.033047}%
  \BibitemOpen
  \bibfield  {author} {\bibinfo {author} {\bibfnamefont {Subhadeep}\
  \bibnamefont {Roy}}, \bibinfo {author} {\bibfnamefont {Soumyajyoti}\
  \bibnamefont {Biswas}}, \ and\ \bibinfo {author} {\bibfnamefont {Purusattam}\
  \bibnamefont {Ray}},\ }\bibfield  {title} {\enquote {\bibinfo {title}
  {Failure time in heterogeneous systems},}\ }\href {\doibase
  10.1103/PhysRevResearch.1.033047} {\bibfield  {journal} {\bibinfo  {journal}
  {Phys. Rev. Res.}\ }\textbf {\bibinfo {volume} {1}},\ \bibinfo {pages}
  {033047} (\bibinfo {year} {2019})}\BibitemShut {NoStop}%
\bibitem [{\citenamefont {Moreira}\ \emph {et~al.}(2012)\citenamefont
  {Moreira}, \citenamefont {Oliveira}, \citenamefont {Hansen}, \citenamefont
  {Ara\'ujo}, \citenamefont {Herrmann},\ and\ \citenamefont
  {Andrade}}]{Moreira2012}%
  \BibitemOpen
  \bibfield  {author} {\bibinfo {author} {\bibfnamefont {A.~A.}\ \bibnamefont
  {Moreira}}, \bibinfo {author} {\bibfnamefont {C.~L.~N.}\ \bibnamefont
  {Oliveira}}, \bibinfo {author} {\bibfnamefont {A.}~\bibnamefont {Hansen}},
  \bibinfo {author} {\bibfnamefont {N.~A.~M.}\ \bibnamefont {Ara\'ujo}},
  \bibinfo {author} {\bibfnamefont {H.~J.}\ \bibnamefont {Herrmann}}, \ and\
  \bibinfo {author} {\bibfnamefont {J.~S.}\ \bibnamefont {Andrade}},\
  }\bibfield  {title} {\enquote {\bibinfo {title} {Fracturing highly disordered
  materials},}\ }\href {\doibase 10.1103/PhysRevLett.109.255701} {\bibfield
  {journal} {\bibinfo  {journal} {Phys. Rev. Lett.}\ }\textbf {\bibinfo
  {volume} {109}},\ \bibinfo {pages} {255701} (\bibinfo {year}
  {2012})}\BibitemShut {NoStop}%
\bibitem [{\citenamefont {Roy}\ \emph {et~al.}(2015)\citenamefont {Roy},
  \citenamefont {Kundu},\ and\ \citenamefont {Manna}}]{Chandreyee2015}%
  \BibitemOpen
  \bibfield  {author} {\bibinfo {author} {\bibfnamefont {Chandreyee}\
  \bibnamefont {Roy}}, \bibinfo {author} {\bibfnamefont {Sumanta}\ \bibnamefont
  {Kundu}}, \ and\ \bibinfo {author} {\bibfnamefont {S.~S.}\ \bibnamefont
  {Manna}},\ }\bibfield  {title} {\enquote {\bibinfo {title} {Fiber bundle
  model with highly disordered breaking thresholds},}\ }\href {\doibase
  10.1103/PhysRevE.91.032103} {\bibfield  {journal} {\bibinfo  {journal} {Phys.
  Rev. E}\ }\textbf {\bibinfo {volume} {91}},\ \bibinfo {pages} {032103}
  (\bibinfo {year} {2015})}\BibitemShut {NoStop}%
\bibitem [{\citenamefont {Hope}\ \emph {et~al.}(2015)\citenamefont {Hope},
  \citenamefont {Kundu}, \citenamefont {Roy}, \citenamefont {Manna},\ and\
  \citenamefont {Hansen}}]{Sigmund2015}%
  \BibitemOpen
  \bibfield  {author} {\bibinfo {author} {\bibfnamefont {Sigmund}\ \bibnamefont
  {Hope}}, \bibinfo {author} {\bibfnamefont {Sumanta}\ \bibnamefont {Kundu}},
  \bibinfo {author} {\bibfnamefont {Chandreyee}\ \bibnamefont {Roy}}, \bibinfo
  {author} {\bibfnamefont {Subhrangshu}\ \bibnamefont {Manna}}, \ and\ \bibinfo
  {author} {\bibfnamefont {Alex}\ \bibnamefont {Hansen}},\ }\bibfield  {title}
  {\enquote {\bibinfo {title} {Network topology of the desert rose},}\ }\href
  {\doibase 10.3389/fphy.2015.00072} {\bibfield  {journal} {\bibinfo  {journal}
  {Frontiers in Physics}\ }\textbf {\bibinfo {volume} {3}} (\bibinfo {year}
  {2015}),\ 10.3389/fphy.2015.00072}\BibitemShut {NoStop}%
\bibitem [{\citenamefont {George~Batrouni}\ and\ \citenamefont
  {Hansen}(1988)}]{Batrouni}%
  \BibitemOpen
  \bibfield  {author} {\bibinfo {author} {\bibfnamefont {Ghassan}\ \bibnamefont
  {George~Batrouni}}\ and\ \bibinfo {author} {\bibfnamefont {Alex}\
  \bibnamefont {Hansen}},\ }\bibfield  {title} {\enquote {\bibinfo {title}
  {Fourier acceleration of iterative processes in disordered systems},}\ }\href
  {\doibase 10.1007/BF01019728} {\bibfield  {journal} {\bibinfo  {journal}
  {Journal of Statistical Physics}\ }\textbf {\bibinfo {volume} {52}},\
  \bibinfo {pages} {747--773} (\bibinfo {year} {1988})}\BibitemShut {NoStop}%
\bibitem [{\citenamefont {Shekhawat}\ \emph {et~al.}(2013)\citenamefont
  {Shekhawat}, \citenamefont {Zapperi},\ and\ \citenamefont
  {Sethna}}]{Shekhawat}%
  \BibitemOpen
  \bibfield  {author} {\bibinfo {author} {\bibfnamefont {Ashivni}\ \bibnamefont
  {Shekhawat}}, \bibinfo {author} {\bibfnamefont {Stefano}\ \bibnamefont
  {Zapperi}}, \ and\ \bibinfo {author} {\bibfnamefont {James~P.}\ \bibnamefont
  {Sethna}},\ }\bibfield  {title} {\enquote {\bibinfo {title} {From damage
  percolation to crack nucleation through finite size criticality},}\ }\href
  {\doibase 10.1103/PhysRevLett.110.185505} {\bibfield  {journal} {\bibinfo
  {journal} {Phys. Rev. Lett.}\ }\textbf {\bibinfo {volume} {110}},\ \bibinfo
  {pages} {185505} (\bibinfo {year} {2013})}\BibitemShut {NoStop}%
\bibitem [{\citenamefont {Hansen}\ and\ \citenamefont
  {Schmittbuhl}(2003)}]{Hansen2003}%
  \BibitemOpen
  \bibfield  {author} {\bibinfo {author} {\bibfnamefont {Alex}\ \bibnamefont
  {Hansen}}\ and\ \bibinfo {author} {\bibfnamefont {Jean}\ \bibnamefont
  {Schmittbuhl}},\ }\bibfield  {title} {\enquote {\bibinfo {title} {Origin of
  the universal roughness exponent of brittle fracture surfaces:stress-weighted
  percolation in the damage zone},}\ }\href {\doibase
  10.1103/PhysRevLett.90.045504} {\bibfield  {journal} {\bibinfo  {journal}
  {Phys. Rev. Lett.}\ }\textbf {\bibinfo {volume} {90}},\ \bibinfo {pages}
  {045504} (\bibinfo {year} {2003})}\BibitemShut {NoStop}%
\bibitem [{\citenamefont {Das}\ and\ \citenamefont
  {Biswas}(2023)}]{soumyaditya}%
  \BibitemOpen
  \bibfield  {author} {\bibinfo {author} {\bibfnamefont {Soumyaditya}\
  \bibnamefont {Das}}\ and\ \bibinfo {author} {\bibfnamefont {Soumyajyoti}\
  \bibnamefont {Biswas}},\ }\bibfield  {title} {\enquote {\bibinfo {title}
  {Critial scaling through gini-index},}\ }\href@noop {} {\bibfield  {journal}
  {\bibinfo  {journal} {arXiv:2211.01281}\ } (\bibinfo {year}
  {2023})}\BibitemShut {NoStop}%
\bibitem [{\citenamefont {Biswas}\ and\ \citenamefont
  {Chakrabarti}(2020)}]{PhysRevE.102.012113}%
  \BibitemOpen
  \bibfield  {author} {\bibinfo {author} {\bibfnamefont {Soumyajyoti}\
  \bibnamefont {Biswas}}\ and\ \bibinfo {author} {\bibfnamefont {Bikas~K.}\
  \bibnamefont {Chakrabarti}},\ }\bibfield  {title} {\enquote {\bibinfo {title}
  {Flory-like statistics of fracture in the fiber bundle model as obtained via
  kolmogorov dispersion for turbulence: A conjecture},}\ }\href {\doibase
  10.1103/PhysRevE.102.012113} {\bibfield  {journal} {\bibinfo  {journal}
  {Phys. Rev. E}\ }\textbf {\bibinfo {volume} {102}},\ \bibinfo {pages}
  {012113} (\bibinfo {year} {2020})}\BibitemShut {NoStop}%
\bibitem [{\citenamefont {Ramstad}\ \emph {et~al.}(2004)\citenamefont
  {Ramstad}, \citenamefont {Bakke}, \citenamefont {Bjelland}, \citenamefont
  {Stranden},\ and\ \citenamefont {Hansen}}]{PhysRevE.70.036123}%
  \BibitemOpen
  \bibfield  {author} {\bibinfo {author} {\bibfnamefont {Thomas}\ \bibnamefont
  {Ramstad}}, \bibinfo {author} {\bibfnamefont {Jan \O{}.~H.}\ \bibnamefont
  {Bakke}}, \bibinfo {author} {\bibfnamefont {Johannes}\ \bibnamefont
  {Bjelland}}, \bibinfo {author} {\bibfnamefont {Torunn}\ \bibnamefont
  {Stranden}}, \ and\ \bibinfo {author} {\bibfnamefont {Alex}\ \bibnamefont
  {Hansen}},\ }\bibfield  {title} {\enquote {\bibinfo {title} {Correlation
  length exponent in the three-dimensional fuse network},}\ }\href {\doibase
  10.1103/PhysRevE.70.036123} {\bibfield  {journal} {\bibinfo  {journal} {Phys.
  Rev. E}\ }\textbf {\bibinfo {volume} {70}},\ \bibinfo {pages} {036123}
  (\bibinfo {year} {2004})}\BibitemShut {NoStop}%
\bibitem{cubic}
https://en.wikipedia.org/wiki/Cubic\_equation
\end{thebibliography}
%

\end{document}